**TIME SERIES FORECASTING: A MULTIVARIATE STOCHASTIC APPROACH**

*Stefano Sello*
*Termo – Fluid Dynamics Research Center*
*Enel Research*
*Via Andrea Pisano, 120*
*56122 PISA - ITALY*

*ABSTRACT*

This note deals with a multivariate stochastic approach to forecast the behaviour of a cyclic time series. Particular attention is devoted to the problem of the prediction of time behaviour of sunspot numbers for the current 23th cycle. The idea is to consider the previous known n cycles as n particular realizations of a given stochastic process. The aim is to predict the future behavior of the current n+1th realization given a portion of the curve and the structure of the previous n realizations. The (linear) model derived is based on the cross-correlations between the current n+1th realization and the previous n ones and the solution of the related least squares problem. As example we applied the method to smoothed monthly sunspots numbers from SIDC archives, in order to predict the behaviour of the current 23th solar cycle.

*1. SHORT DESCRIPTION OF THE METHOD*

A cyclic time series, independently of the degree of complexity, may be partizioned in n subsets according to its main cycle. These subsets are considered as n different realizations of a given stochastic process. In this approach the problem of forecasting a part of a next cycle is equivalent to estrapolate a reliable behaviour for the n+1th stochastic realization from the known structure of the previous n realizations. The criteria adopted here is to relate the known part of the n+1th realization with the equivalent part of the n realizations. This allows the set up of a predictive model for the unknown part of the n+1th realization. More precisely, the method starts with the computation of the cross-correlation functions for all the known couple of realizations: (yn+1,yk), k=1,2,…,n: [1]

$$R_{yn+1, yk}(\tau) = \lim_{t \to \infty} \frac{1}{T} \int_0^T y_{n+1}(t) y_k(t+\tau) dt$$

This allows to derive 1) the degree of correlation of the known part of the n+1th realization with the previous n realizations and 2) the related optimal lag with respect to the known part of the n+1th realization. This information is necessary to set up an optimal predictive model.

The linear model derived from the correlation analysis is of the form:

$$y_{n+1}(t) = c_0 + \sum_{i=1}^{s} c_i y_i(t+\tau_i)$$

where s is the number of known realizations with a degree of cross-correlation greater than a defined value. The set of coefficients $c_k$ is evaluated through a least squares problem using the known values of the n+1th realization.
This model is then utilized to predict the future behaviour of the n+1th realization.

2. APPLICATION TO THE SOLAR CYCLE: THE SMOOTHED MONTHLY SUNSPOTS NUMBERS

As application of the above method we considered the well known solar cycle through the time series of the smoothed monthly sunspots numbers (ssnm) from the Sunspot Index Data Center (SIDC) of Brussels, Belgium [2]. As we know the about 11 years solar cycle is evidenced in the whole complexity by the time series of the sunspots numbers. The behaviour is charaterized by a very noisy data with high irregular and dispersive cycles. From a statistical viewpoint the 22 cycles known are non-homogeneous and non-stationary. The accuracy of any predictive model is then strongly affected by these characteristics. Figure 1 shows the sequential time series of the ssnm with the related wavelet analysis in order to point out the principal set of global and local oscillations inside the data. This analysis allows a precise determination of the shape and extension of the main oscillation.

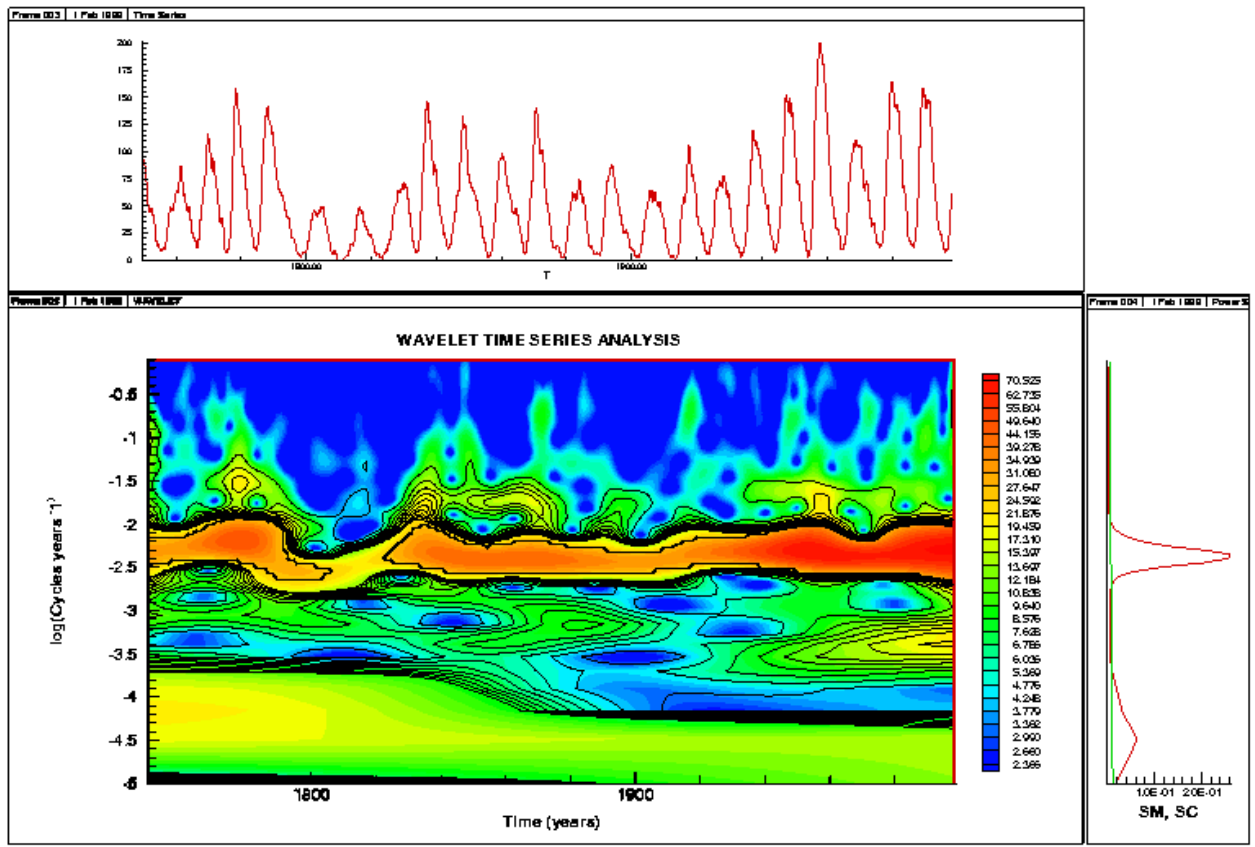

Figure 1

The n=22 cycles are then considered as known realizations of the underlying stochastic process. At the present (01/01/1999) we know only the first 26 values of the ssnm for the current 23th cycle.  These values jointed with the known structure of the previous cycles allow the set-up of a predictive model  based on cross-correlations functions. The basic idea is to handle each cycle as a particular realization of the same stochastic process. The predictive model is the result of the cross-correlations existing between the known part of the 23th cycle and  the remaining cycles. As known, the time extension of the 22 cycles is different and thus we have limited the analysis to the first common 100 values from the evaluated minimum (dt=0.082 years). The result of the correlation analysis was that only s=8 cycles are strongly correlated (R≥0.8) with the known part of the current 23th cycle: k=2,9,10,11,12,13,14,22. Thus the model is formed by a proper linear combination of the related cycles. Next a least squares problem was solved in order to determine the values of the unknown s+1 coefficients. The maximum relative error was about 10%. The model is then used to predict the behaviour of the successive values of ssnm for the 23th cycle.

In Figure 2 we show the results. In particular, as indicated, the current prediction, based on this approach, is that the next ssnm peak will reach the value: 147.76 at 1999.89. A number of computations on known cycles showed that the tendency of the method is to overestimate the peak. We stress the comparison of the predicted cycle shape with the previous cycles. The high irregular and dispersive level of the previous n realizations (both in phase and in intensity) forces the predicted behaviour to be very unstable and subjected to significative variations. Of course, as new values will be included in the current cycle, the analysis should imply more and more reliable predictions.

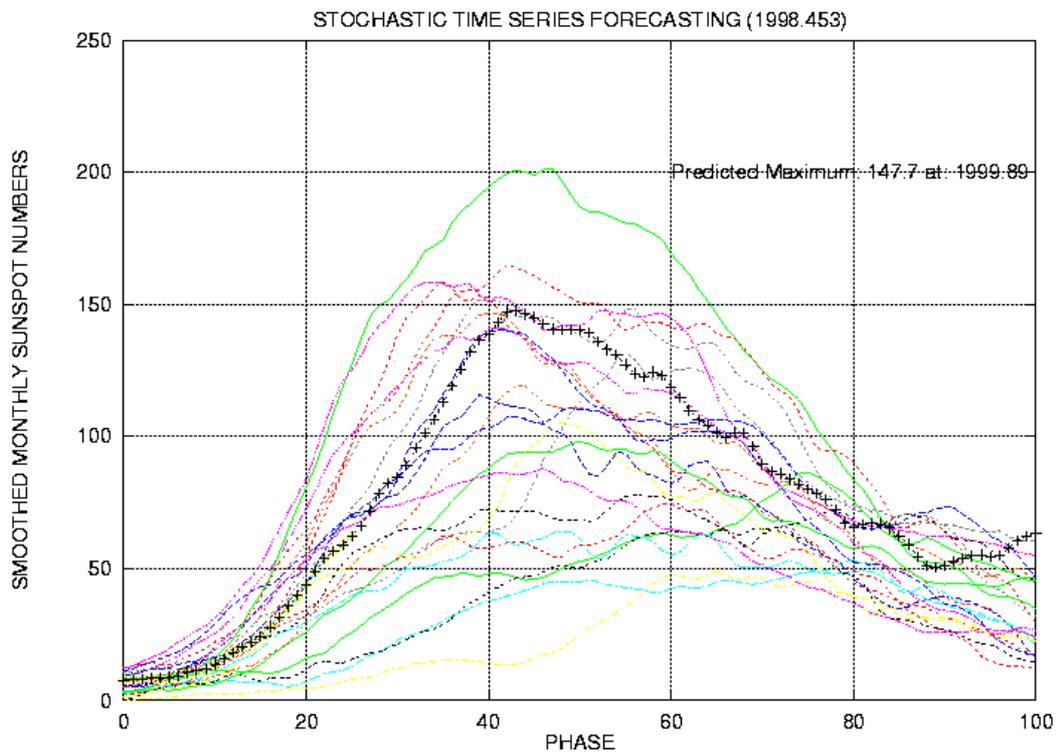

Figure 2

CONCLUSIONS

The above analysis shows a possible method to set-up a predictive model of cyclic time series based on a multivariate stochastic approach. Particular attention is devoted to the problem of the prediction of time behaviour of sunspot numbers for the current 23th cycle. The basic idea is to consider the previous known n cycles as n particular realizations of a given stochastic process. The aim is to predict the future behavior of the current n+1th realization given a portion of the curve and the structure of the previous n realizations. The quantitative results are consistent with other predictive methods such as based on correlations with geomagnetic indices, rarm models, etc.

*References*

[1] J.S. Bendat, A.G. Piersol: "Random Data: Analysis and Measurement Procedures", Wiley-Interscience Ed. (1971)

[2] SIDC Archives: http://www.oma.be/KSB-ORB/SIDC/index.html

*APPENDIX*

As a comparison Figure 3 shows a computation from SIDC with daily sunspot number (yellow), monthly mean sunspot number (blue), smoothed monthly sunspot number (red) and predictions of the smoothed monthly sunspot number for the last eleven years, up to now. SM (red dots): classical prediction method, based on an interpolation of Waldmeier standard curves. CM (red dashes): combined method (K. Denkmayr), a regression technique coupling a dynamo-based estimator with Waldmeier idea of standard curves.

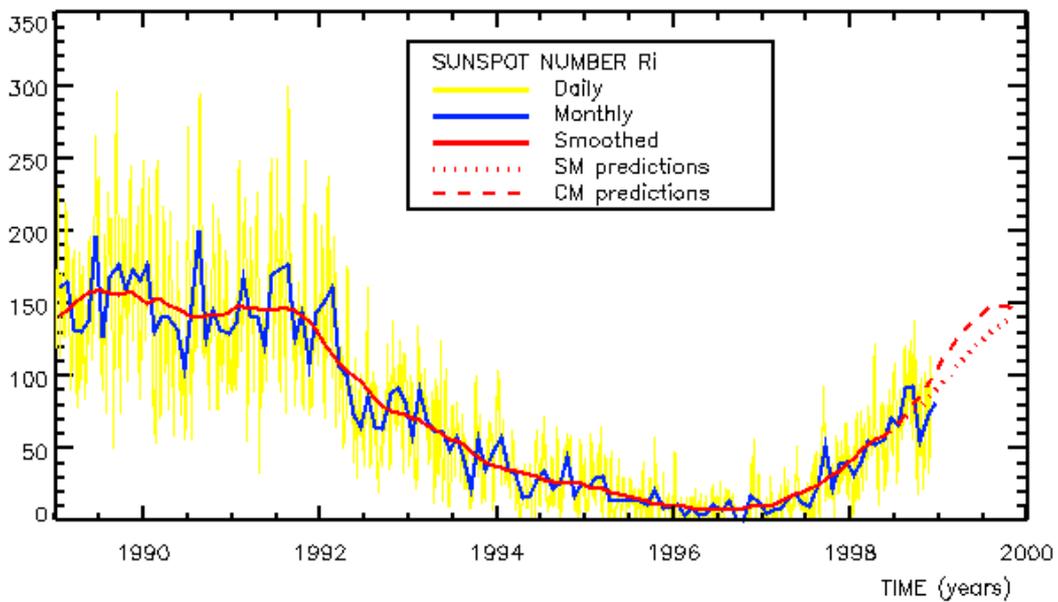

Figure 3